\documentclass[preprint,12pt]{elsarticle}
\usepackage{amsmath}
  \usepackage{graphics} %% add this and next lines if pictures should be in esp format
\usepackage{graphicx}  %% add this and next lines if pictures should be in esp format
  \usepackage{epsfig} %For pictures: screened artwork should be set up with an 85 or 100 line screen
  \usepackage{float}
  \usepackage{caption}
\usepackage{amsmath}
\usepackage{wrapfig}
\usepackage{graphicx}
\usepackage{subcaption}
\usepackage{amssymb}
\usepackage{caption}

\usepackage{color}
\usepackage{enumitem}
\usepackage{comment}

\newcommand{\beq}{\begin{eqnarray*}}
\newcommand{\eeq}{\end{eqnarray*}}
\def\eqref#1{$(\ref{#1})$}
\newcommand{\lp}{\left(}
\newcommand{\rp}{\right)}
\newcommand{\at}{\alpha}
\newcommand{\RR}{r}

\begin{document}

\begin{frontmatter}

\title{Optimal Control and Analysis of a Modified Trojan Y-Chromosome Strategy}
\author{Matthew A. Beauregard$^{1}$, Rana D. Parshad$^{2}$, Sarah Boon$^{1}$, Harley Conaway$^{1}$, Thomas Griffin$^{1}$, Jingjing Lyu$^{3}$ }
\address{$^1$Department of Mathematics and Statistics, Stephen F. Austin State University, Nacogdoches, TX, 75962, USA
\\
$^2$Department of Mathematics, Iowa State University, Ames, IA, 50011, USA
\\
$^3$Department of Mathematics, DePaul University, Chicago, IL, 60604, USA
}

\begin{abstract}
The Trojan Y Chromosome (TYC) Strategy is a promising eradication method that attempts to manipulate the female to male ratio to promote the reduction of the population of an invasive species.  The manipulation stems from an introduction of sex-reversed males, called supermales, into an ecosystem.  The offspring of the supermales is guaranteed to be male. Mathematical models have shown that the population can be driven to extinction with a continuous supply of supermales.  In this paper, a new model of the TYC strategy is introduced and analyzed that includes two important modeling characteristics, that are neglected in all previous models.  First, the new model includes intraspecies competition for mates.  Second, a strong Allee effect is included.  Several conclusions about the strategy via optimal control are established. These results have large scale implications for the biological control of invasive species.
\end{abstract}


%\subjclass{Primary: 34C11, 34C23, 49J15; Secondary: 92D25, 92D40}
\end{frontmatter}

%%%%%%%%%%%%%%%%%%%%%%%%%%%%%%%%%%%%%%%%%%%%%%%%%%%%%%%%%%%%%%%%%%%%%%%%%%%%%%%%%%%%%%%%%%%%%%%%%%%%%%%%%%%%%%%%%%%%%%%%%%%%%%%%%
%%%%%%%%%%%%%%%%%%%%%%%%%%%%%%%%%%%%%%%%%%%%%%%%%%%%%%%%%%%%%%%%%%%%%%%%%%%%%%%%%%%%%%%%%%%%%%%%%%%%%%%%%%%%%%%%%%%%%%%%%%%%%%%%%
%%%%%%%%%%%%%%%%%%%%%%%%%%%%%%%%%%%%%%%%%%%%%%%%%%%%%%%%%%%%%%%%%%%%%%%%%%%%%%%%%%%%%%%%%%%%%%%%%%%%%%%%%%%%%%%%%%%%%%%%%%%%%%%%%

\section{Introduction}

The detrimental effects of aquatic invasive species is well-documented \cite{A06, A12, B07, C01, L12, M00, 089, S97, V96}.  Subsequently, a tremendous amount of effort by habitat controllers is devoted to designing effective eradication strategies such as chemical treatment, local harvesting,  dewatering, ichthyocides, or a suitable combination \cite{SL15}. However, these methods are known to negatively impact ecosystems, which may be already stressed by the presence of an aquatic invasive species \cite{SL15}.

\begin{figure}[H]
 \centering{
  {\includegraphics[scale=.75]{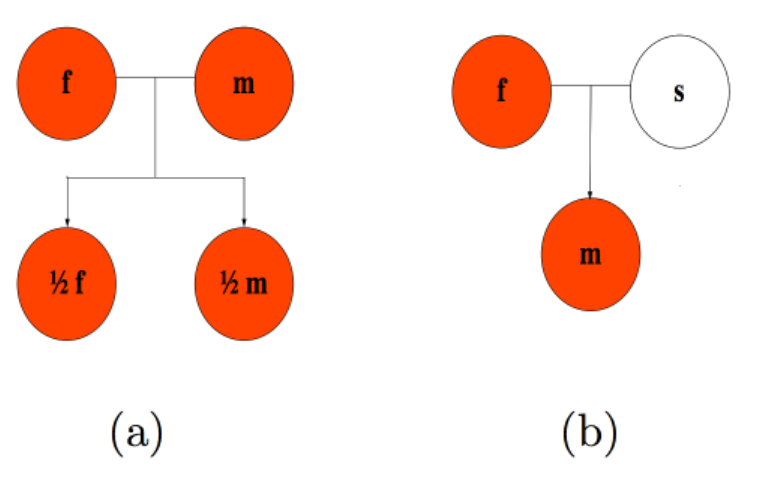}
  }
 \caption{\small The pedigree tree of the TYC model (that demonstrates {\it Trojan Y-Chromosome} eradication strategy).
 (a) Mating of a wild-type XX female (f) and a wild-type XY male (m). (b) Mating of a wild-type XX female (f) and a YY supermale (s). Red color represents wild types, and white color represents phenotypes.}
 \label{Fig:TYC-Eradication-Strategy}
 }
 \end{figure}

The Trojan Y chromosome strategy (TYC) is a new eradication strategy which circumvents many of the known negative ecological impacts due to current practice \cite{GutierrezTeem06, TGP13, SL15}.  TYC strategy involves an introduction of a sex-reversed male.  The off-spring of the sex-reversed male, called a supermale, with a wild-type female is guaranteed to be male.  Therefore, subsequent generations become male-dominant and this skews the sex ratio towards more males. The goal, is that through the gradual reduction in the female population, extinction of the population may occur (see Fig.~\ref{Fig:TYC-Eradication-Strategy}).  The supermale is not a genetically modified organism (GMO) and the TYC process is reversible, that is, if the introduction of supermales is stopped then the supermale population will die out \cite{Schill2017}. The TYC strategy has seen tremendous experimental and theoretical interest \cite{Schill18,Schill16,TGP13, CW07a, CW07b, CW09, Gutierrez12, P11, p10, ParshadGutierrez11, Parshad13, P09,  SDE2013, ODE2013, Z12}.

The classical population model of the TYC strategy relates the populations of the wild-type XX females ($f$), wild-type XY males ($m$), and the YY supermale ($s$) populations over time.  A mathematical model was first proposed by Gutierrez, Teem, and Parshad in \cite{GutierrezTeem06, TGP13}:

\begin{eqnarray}
\label{ClassicTYCfeq} \dot{f} &=& \frac{1}{2}  \beta L fm - \delta f,\\
\label{ClassicTYCmeq} \dot{m} &=& \frac{1}{2}  \beta L fm + \beta L fs- \delta m,\\
\label{ClassicTYCseq} \dot{s} &=& \mu - \delta s,
\end{eqnarray}
where $L = 1 - \frac{f+m+s}{K}$, $K$ is the carrying capacity, $\beta$ is the birth rate, $\delta$ is the death rate, and $\mu$ is the constant introduction rate.  The parameters and populations are assumed to be nonnegative.  Due to the nonlinearities, it is not necessary to assume that $\beta > \delta>0$, to obtain a persistent invasive population. It was shown in \cite{TGP13} that there exists a $\mu^*$ such that for all $\mu > \mu^*>0$ that $f,m \to 0$ in infinite time.

In the current manuscript, we seek to include two important and relevant modeling features.  First, the influx of supermales leads to competition between wild-males and supermales for female mates. Second, if the female population is below a given threshold then the population loses fitness and extinction should occur \cite{Drake2011,Kramer2009}.  This latter phenomena is called the Allee effect.  These two modeling features are introduced in Section 2.  A stability analysis of the equilibrium solutions is also included in Section 2. In Section 3, we then investigate the influence that intraspecies competition and the Allee effect have on the optimal introduction rate that minimizes an objective function based on the total wild population and introduced super males.  A stochastic model is introduced to examine the influence of noise on the birth and death rates have on the objective function's value at the optimal introduction rate.

%%%%%%%%%%%%%%%%%%%%%%%%%%%%%%%%%%%%%%%%%%%

\section{Modified TYC Model with Strong Allee Effect}

In this paper, we investigate and propose a new model of the TYC strategy.  Namely,
\begin{eqnarray}
\label{feq} \dot{f} &=& \frac{1}{2}  \beta L \lp \frac{f}{\at}-1 \rp \rho_1(m,s) fm - \delta f,\\
\label{meq} \dot{m} &=& \frac{1}{2}  \beta L \lp \frac{f}{\at}-1 \rp \rho_1(m,s) fm + \beta L \lp \frac{f}{\at}-1 \rp  \rho_2(m,s) fs- \delta m,\\
\label{seq} \dot{s} &=& \mu - \delta s,
\end{eqnarray}
where $L, \mu, \beta,$ and $\delta$ are as before. Again, the parameters and populations are assumed to be nonnegative and that $\beta > \delta>0$, that is, the birth supercedes the death rate.

Intraspecies competition between wild male and supermale populations for female mates is modeled through the nonnegative saturation term:
\beq
\label{rho12} \rho_1(m,s)=\frac{m}{m+s}~,~~~ \rho_2(m,s) =\frac{s}{m+s}.
\eeq
The saturation terms provide the percentage of the total male population that is either wild-type or supermale.  Clearly, the range of the saturations term is $[0,1].$ Notice, for a fixed wild-type population, as $s \rightarrow \infty$ then $\rho_1\rightarrow 0^+$ and $\rho_2\rightarrow 1^-$.  In this situation, the birth of females approaches zero while male offspring only occurs from female and supermale progeny.  Likewise, for a fixed supermale population, as $m \rightarrow \infty$ then $\rho_1\rightarrow 1^-$ and $\rho_2\rightarrow 0^+$; subsequently, male progeny only occurs from wild-type male and female mating.  Therefore, $\rho_1$ and $\rho_2$ attempt to model the difficulty of wild-type female finding suitable mates from either wild-type or supermale populations.

The term $f/\at - 1$ models a strong Allee effect and represents a loss of fitness in the female population when below the Allee threshold, $\at$.  This effect is problem of \textit{undercrowding} of a species and was first motivated by observations made by Allee in 1927 \cite{Allee1927, Allee1931, Allee_1954} . Since then, numerous evidences of this effect have been established \cite{Kramer2009,Stephens_1999}.  Notice, that when $f/\at - 1<0$ then $\dot{f}<0$, provided $L>0$.  Therefore, the female population will decrease toward extinction.  Subsequently, a goal of the TYC strategy is to \textit{push}, via the introduction of supermales, the female population below the Allee threshold.

In the forthcoming analysis, the equations are rescaled.  The populations are scaled by the carrying capacity while the $t$ is scaled by the deathrate, that is $t\rightarrow t/\delta$.  The rescaled equations are:
\begin{eqnarray}
\label{feq1} \dot{f} &=& \RR L \lp \frac{f}{a}-1 \rp \lp \frac{m}{m+s} \rp fm - f,\\
\label{meq1} \dot{m} &=& \RR \frac{Lf}{m+s} \lp \frac{f}{a}-1 \rp \lp m^2 + 2 s^2 \rp - m,\\
\label{seq1} \dot{s} &=& \gamma - s,
\end{eqnarray}
where $L \rightarrow 1 - (f+m+s)$, $\RR = \dfrac{K\beta}{2\delta}>1$, $a = \dfrac{\at}{K}\ll 1$, and $\gamma = \dfrac{\mu}{K \delta}.$

\subsection{Equilibria and Stability Analysis.}

A clear requirement of a valid intervention strategy is that in the absence of the strategy the invasive population would persist, while if the intervention strategy was employed control, ideally extinction, of the invasive species would be established.   Therefore, an effective TYC strategy is one that provides an introduction rate $\gamma$ and initial supermale population such that the wild-population is driven to extinction, while in the supermale free case the population would persist.

It is clear that in the modified model exhibits the equilibrium solution $(0,0,\gamma)$, which is deemed the \textit{extinction state}.  Once $f < a$, the introduction rate of supermales, $\gamma$, should be set zero, causing $s \to 0$.  Subsequently, we investigate the equilibrium solutions in the case where $\gamma = 0$.  A necessary requirement for a successful TYC strategy that there exists some time such that In the forthcoming equilibrium analysis we are interested in equilibrium solutions in the situation where $\gamma = 0$.  Obviously, when $\gamma=0$ then in equilibrium $s=0$, therefore the intraspecies terms $\rho_1(m,0) = 1$ and $\rho_2(m,0) = 0.$  In such case:

\[ f\left( r L\left( \dfrac{f}{a} - 1\right) m - 1 \right) = 0 ~~~ m\left( r L\left( \dfrac{f}{a} - 1\right) f - 1 \right) = 0 \]

We seek to analyze the stability and presence of nontrivial equilibrium solutions.  Noting that in equilibrium $f/a - 1 \neq 0$ then

\[ m = f = \dfrac{1}{rL(f/a - 1)} \]

Hence, all equilibrium solutions fall on the line $f=m$.  The nontrivial equilibrium solutions are roots to the third degree polynomial:
\begin{equation}
\label{EquilEq} g(f)=-2 f^3 +(2a + 1)f^2 - a f -\frac{a}{r}.
\end{equation}
By Descartes rule of signs there always exists a one negative real root, which is neglected since this is not realistic.  In addition, there are either two or zero positive real roots.  If there are zero positive real roots then the only equilibrium solution is the extinction state and is globally attracting. In such case, the TYC strategy is not necessary.  Therefore, we assume there are two real roots, $f_1$ and $f_2$, where $0<f_1<f_2.$

The Jacobian of our system of equations is
\begin{eqnarray*}
 J = \left( \begin{array}{ccc} \kappa + \mu - 1 & \kappa     & c \\
                               \kappa + \mu     & \kappa - 1 & c  \\
                               0 & 0   & -1
             \end{array}\right)
\end{eqnarray*}
where
\beq
\kappa &=&  (1-2 f) f \left(-1+\frac{f}{a}\right) r-f^2 \left(-1+\frac{f}{a}\right) r \\
c      &=& -(1-2 f) f \left(-1+\frac{f}{a}\right) r-f^2 \left(-1+\frac{f}{a}\right) r \\
\mu &=& \frac{(1-2 f) f^2 r}{a}
\eeq
Clearly, $\lambda_3=-1$ is an eigenvalue, which indicates exponentially decay in the supermale population.  The remaining two eigenvalues are determined by the characteristic equation of the submatrix $J_{33}$. Namely,
\beq
\lambda_{1,2} &=& \frac{\mbox{tr}(J_{33}) \pm \sqrt{ \mbox{tr}(J_{33})^2 - 4 \mbox{det}(J_{33})}}{2} \\
&=& -1,~ 2\kappa + \mu - 1.%\frac{-a-2 a f r+3 f^2 r+6 a f^2 r-8 f^3 r}{a}
\eeq
To investigate the sign of $\lambda_2$ we recall that $f_i$ is a root of Eq.~\eqref{EquilEq}.  This fact, is used to determine an expression for $\lambda_2$ as a function of $f_i$:

\[ \lambda_2(f_i) = 3 + 2f_i r - \left(2 + \frac{1}{a}\right)f_i^2 r \]

Notice, that this function is a concave down quadratic with a maximum location located at $f = 1/(2+1/a) > 0$.  Define $f_+$ is the positive root of the quadratic function $\lambda_2(f)$. %$f_+ = \dfrac{a r+\sqrt{3 a r+6 a^2 r+a^2 r^2}}{r+2 a r}$
Then $\lambda_2>0$ if $f_i<f_+$ and $\lambda_2<0$ if $f_i>f_+.$  Notice that $\lambda_2(1) = 3 - r/a$.  Since $a\ll 1$ implies that $r/a \gg 1$ then $\lambda_2(1) < 0$.  Subsequently, we have $0<f_+<1$.

Notice that $g(0), g(1)<0$.  Now, $g(f_+)>0$ provided that

\begin{eqnarray*}
& &q(r,a) = r \left(2+16 a^3+a r+4 a^3 r+\sqrt{a r (3+6 a+a r)}+4 a^2 \sqrt{a r (3+6 a+a r)}\right)\\
& &~~ - 2 (3+6 a+2 a r) \left(a r+\sqrt{a r (3+6 a+a r)}\right) > 0 \\
%& &2+16 a^3+a r+4 a^3 r+\sqrt{a r (3+6 a+a r)}+4 a^2 \sqrt{a r (3+6 a+a r)} \\
%& &~~ > 2 (3+6 a+2 a r) \left(a r+\sqrt{a r (3+6 a+a r)}\right)
\end{eqnarray*}

\begin{figure}[H]
  \includegraphics[scale=.4]{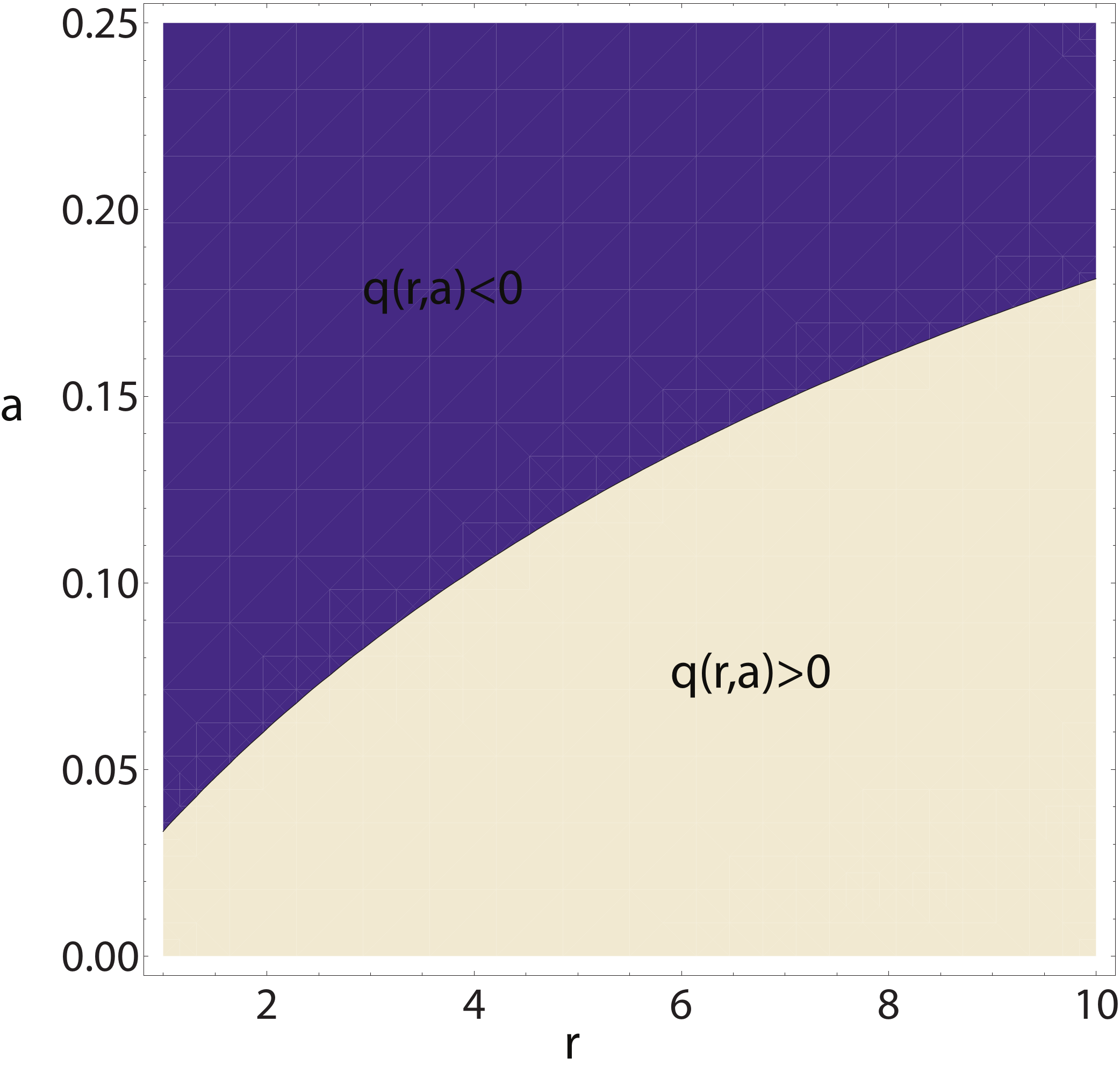}
  \caption{Plot of $q(r,a)$ indicating regions where $q(r,a)$ is greater than zero and less than zero. }
  \label{contourofq}
\end{figure}

Fig.~\ref{contourofq} shows the region where $q(r,a)>0$ and $q(r,a)<0$.  While the Allee threshold is difficult to know precisely for any biological system \cite{Drake2011}, it is reasonable to expect values less than $5\%$ of the carrying capacity \cite{Drake2011,Allee_1954}.  In such case,  $g(f_+)>0$ then by the Intermediate Value Theorem we have $f_1<f_+<f_2$.  Subsequently, $(f_1,f_1,0)$ and $(f_2,f_2,0)$ are a saddle and a sink, respectively.  We call $(f_2, f_2,0)$ the \textit{sustained state}.  An effective TYC strategy will push populations away from the sustained state and toward the basin of attraction, ideally into, of the extinction state.

\section{Optimal Control Analysis}
\subsection{Optimal control analysis}
The goal of this section is to investigate the mechanisms in our TYC system of equations, that, if controlled, could lead to optimal levels of both densities. We assume that the scaled introduction rate $\gamma$ is not known \emph{a priori} and enter the system as a time-dependent control, such that $0 \leq \gamma(t) < \infty$. Consider the objective function
\beq%\begin{equation}
J_0(\gamma)=\int^{T}_{0} -(f+m)- \frac{1}{2}\gamma^2 dt
\eeq%\end{equation}
%
% View OptControlIntraSpecies.m using Tomlab
%    In simulations an unscaled objective function is used.
%
subject to the governing equations \eqref{feq1}-\eqref{seq1} and initial conditions. Optimal strategies are derived for the objective function, where we minimize both female and male populations while also minimizing the introduction rate $\gamma$. Optimal controls are searched for within the set $U_0$, namely,
\beq%\begin{equation}
 U_0= \{ \gamma~|~\gamma \ \mbox{measurable}, \  0 \leq \gamma < \infty,  \  t \in [0,T], \ \forall T\}.
\eeq%\end{equation}
The goal is to seek an optimal $\gamma^{*}(t)$ such that,
\begin{equation}
\label{objectivefunc}
J_0(\gamma^*) =\underset{\gamma} {\max} \displaystyle\int^{T}_{0} -(f+m)- \frac{1}{2}\gamma^2 dt
\end{equation}

We use the Pontryagin's maximum principle to derive the necessary conditions on the optimal control \cite{Lenhart_OptControl}. The Hamiltonian for $J_0$ is given by
\beq%\begin{equation}
\label{TYC_1_OCT_ham}
H_0=-(f+m) - \frac{1}{2}\gamma^2+\lambda_1 f'+\lambda_2 m'+\lambda_3 s'
\eeq%\end{equation}
We use the Hamiltonian to find a differential equation of the adjoint $\lambda_i, i=1,2,3.$ Namely,
\begin{equation*}
\begin{split}
\lambda_1'(t) =&\lambda_2\dfrac{r(m^2+2s^2)}{m+s}\left(f\left(\frac{f}{a}-1\right) + \left(\frac{2f}{a}-1\right)(f+m+s-1)\right)+\\
& \lambda_1 \dfrac{m^2r}{m+s} \left(f\left(\frac{f}{a}-1\right) + \left(\frac{2f}{a}-1\right)(f+m+s-1)+1\right)+1\\
\lambda_2'(t)=& \left(\frac{f}{a}-1\right)\left[ \lambda_1 \dfrac{fmr}{m+s}\left(m+2(f+m+s-1)-\frac{m(f+m+s-1)}{m+s}\right)\right.+\\
 & \left. \lambda_2 \dfrac{fr(m^2+2s^2)}{m+s} + \lambda_2\frac{fr(f+m+s-1)}{m+s}\left(2m - \frac{m^2+2s^2}{m+s}\right)\right] +1 \\
\lambda_3'(t)=& \left(\frac{f}{a}-1\right)\frac{fr}{m+s}\left[ \lambda_2\left\{(f+m+s-1)\left(4s-\frac{(m^2+2s^2)}{m+s}\right) + \right. \right. \\
& \left. \left. (m^2+2s^2) \frac{}{}\right\}+ \lambda_1\left(m^2-\frac{m^2(f+m+s-1)}{(m+s)}\right)\right]+\lambda_3
\end{split}
\end{equation*}
with the transversality condition given by
\begin{equation*}
\lambda_1(T)=\lambda_2(T)=\lambda_3(T)=0
\end{equation*}

In consideration of the optimality conditions, the Hamiltonian function is differentiated with respect to control variable $\gamma$ resulting in:
\begin{equation*}
\frac{\partial H}{\partial \gamma}=\lambda_3 - \gamma
\end{equation*}
A compact way of writing the optimal control $\gamma^*$ is
\begin{equation*}
\gamma^*(t)=\max{(0,\lambda_3)}
\end{equation*}

\subsection{Numerical simulations}
\begin{figure}[H]
 \centering{
  \includegraphics[scale=0.45]{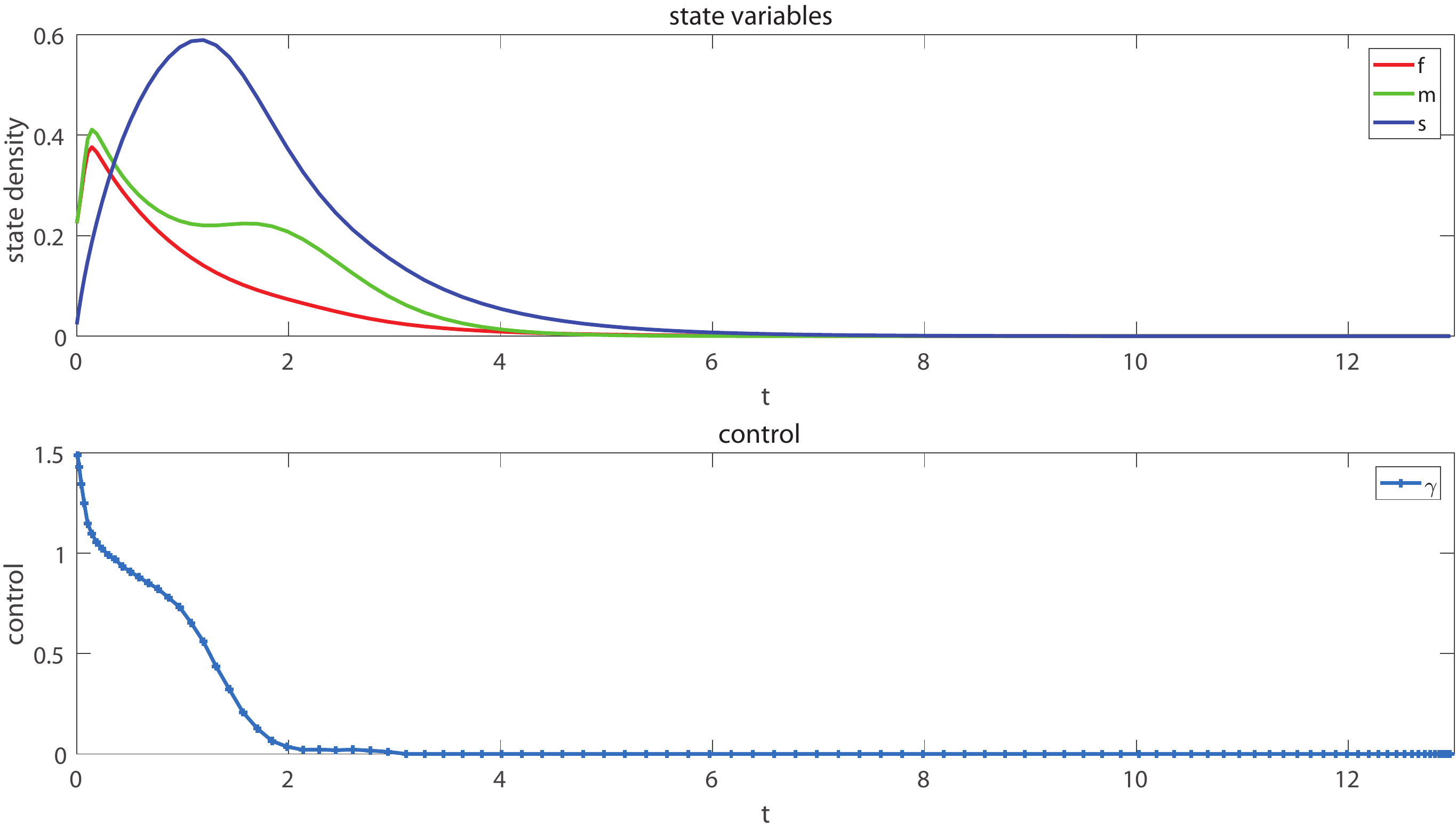}
  }
 \caption{Female (top-red), male (top-green) and supermale (top-blue) densities and optimal control on $\gamma(t)$ (bottom) in change with time $t$. for the modified equations \eqref{feq1}-\eqref{seq1}.
 \label{Fig:opt2}
 }
 \end{figure}

In this section, we will numerically simulate the optimal control for the modified TYC model with strong Allee effect. The following unscaled parameters used for simulation are provided from a least squares approximation of population experiments of guppy fish in \cite{JingJing2019}, namely, $\beta=0.0057, \delta=0.0648, K=405, \at = 24$, and time interval of $(0,200)$. In scaled variables:
\[ r \approx 17.8125, a \approx .06 , 0 < t < 12.96.\]
In Fig.~\ref{Fig:opt2}, the numerically determined optimal $\gamma(t)$ is shown in conjunction with the scaled populations using the optimal control $\gamma(t)$.  This indicates, that a graduate introduction of supermales is ideal.  Notice that the maximum introduction rate is located near the inflection point of the female population at $t\approx 4.25$.  Therefore, the acceleration of the declining population can be used as an indicator as to when to begin rapidly reducing the introduction rate.
\begin{figure}[ht]
 \centering{
\includegraphics[scale=0.45]{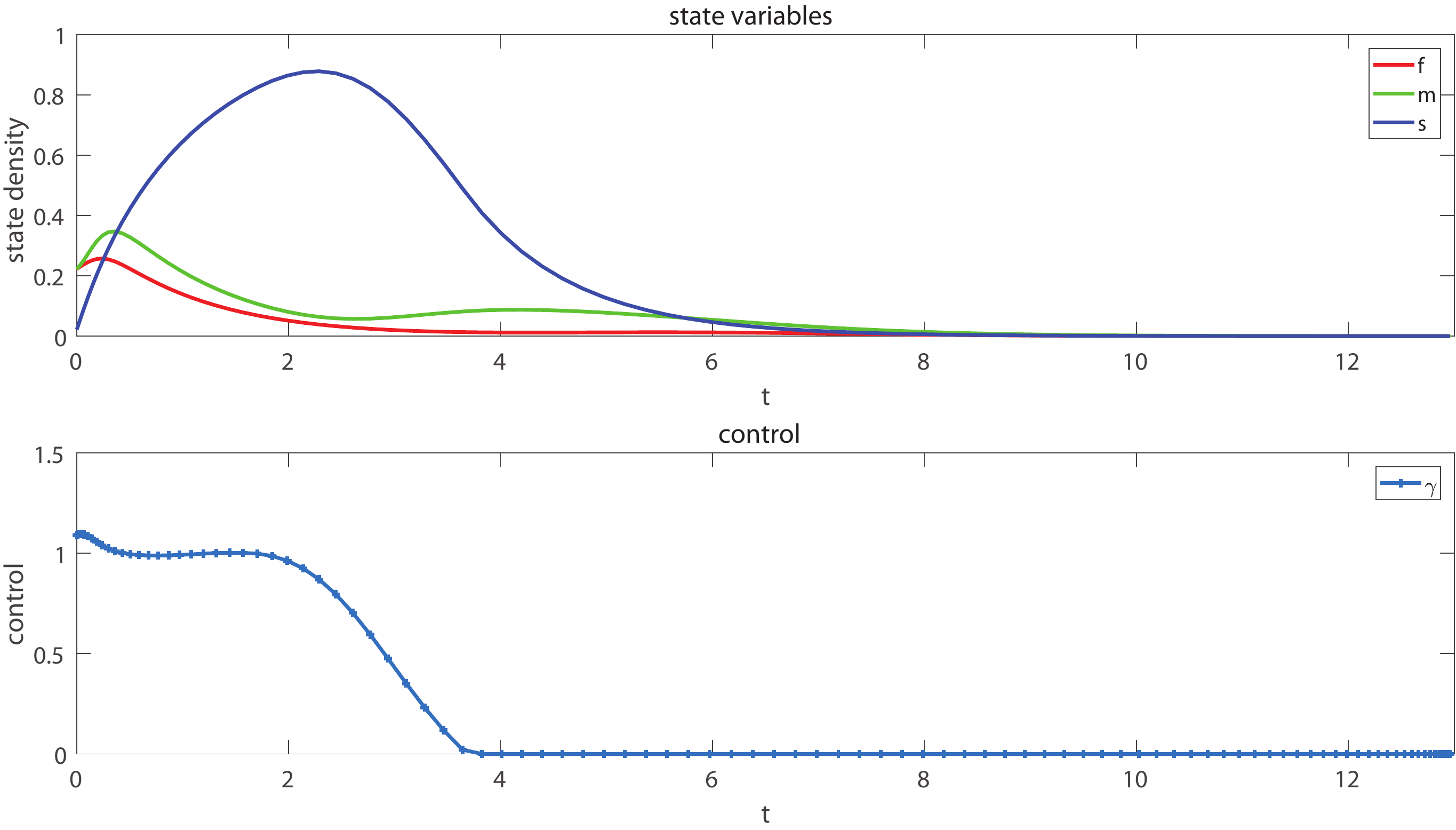}
  }
 \caption{Female (top-red), male (top-green) and supermale (top-blue) densities and optimal control on $\gamma(t)$ (bottom) in change with time $t$ for the classical TYC model
 \label{Fig:optClasical}
 }
 \end{figure}
As a basis of comparison, the optimal $\mu$ in unscaled variables for the classical model was determined in \cite{JingJing2019} (see Fig.~5), which is shown in Fig.~\ref{Fig:optClasical} in scaled variables for convenience.  Without modeling intraspecies competition and the Allee effect yields a larger objective function value at the optimal introduction rate.  In particular, the objective values for the classical and modified models are $-501.9203$ and $-315.1675$, respectively.  The reduced value of the objective function in the modified model is clear by inspection of the plots of the integrands of the corresponding objective functions given in Fig.~\ref{Fig:objfuncs}. Therefore, the inclusion of intraspecies competition and the Allee effect greatly influence the overall cost of the TYC strategy and indicate the strategy is less costly than previously considered in \cite{JingJing2019}.  In the case, of the TYC model that includes intraspecies competition for mates and does not consider the Allee effect obtains an objective value of $-381.3074$, which is still considerably lower than the classical model. However, this suggests that intraspecies competition for mates has a greater influence on the choice of $\gamma$ than the Allee effect.

 \begin{figure}[H]
 \centering{
  \includegraphics[scale=0.45]{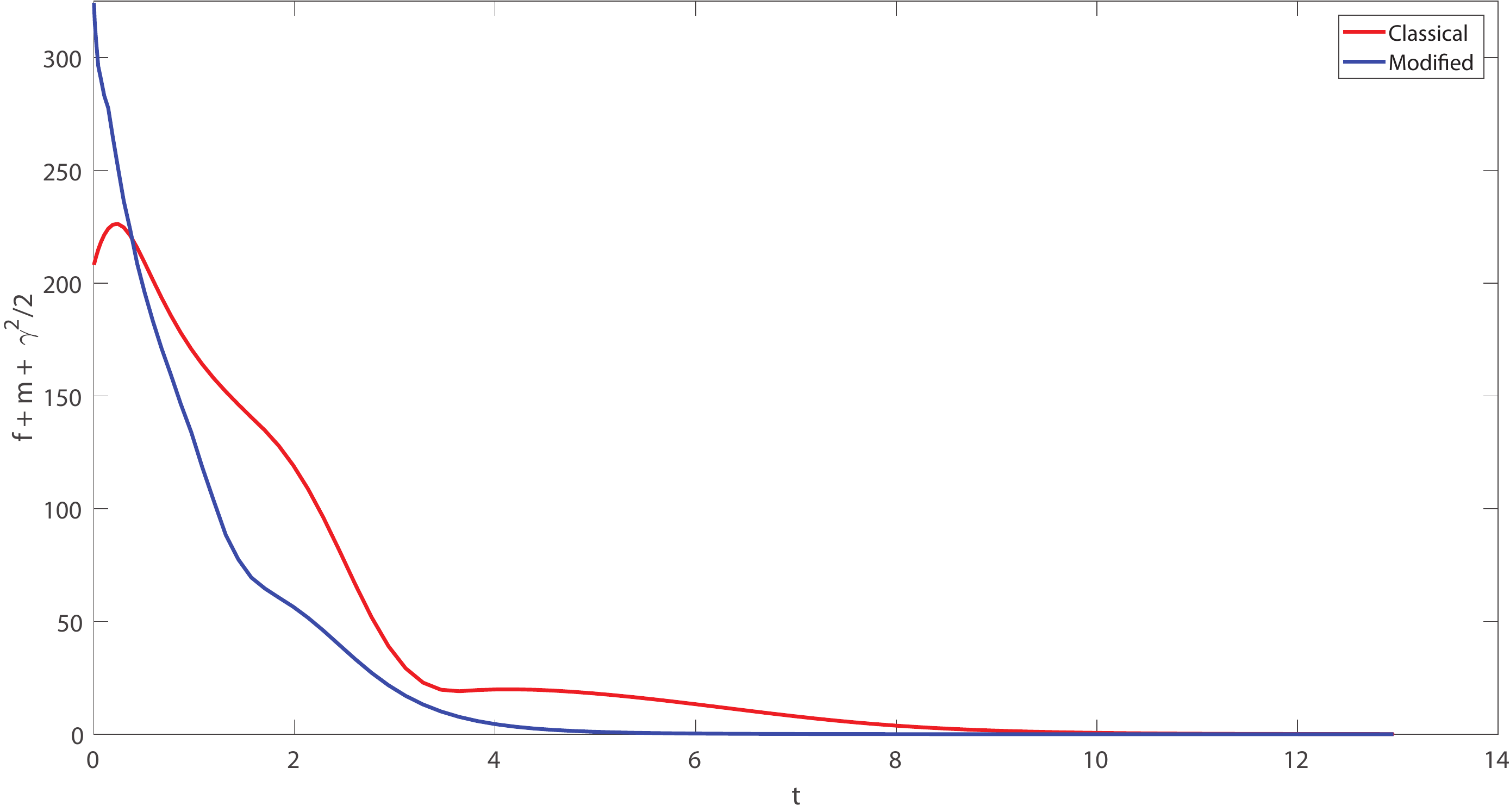}
  }
 \caption{A plot of the integrands of the objective functions, $f+m+\dfrac12 \gamma^2$, for the classical (red) and modified (blue) models.  We notice that the area under the curve for the modified model is less than that of the classical case.
 \label{Fig:objfuncs}
 }
 \end{figure}

\begin{figure}[ht]
 \centering{
  \includegraphics[scale=0.45]{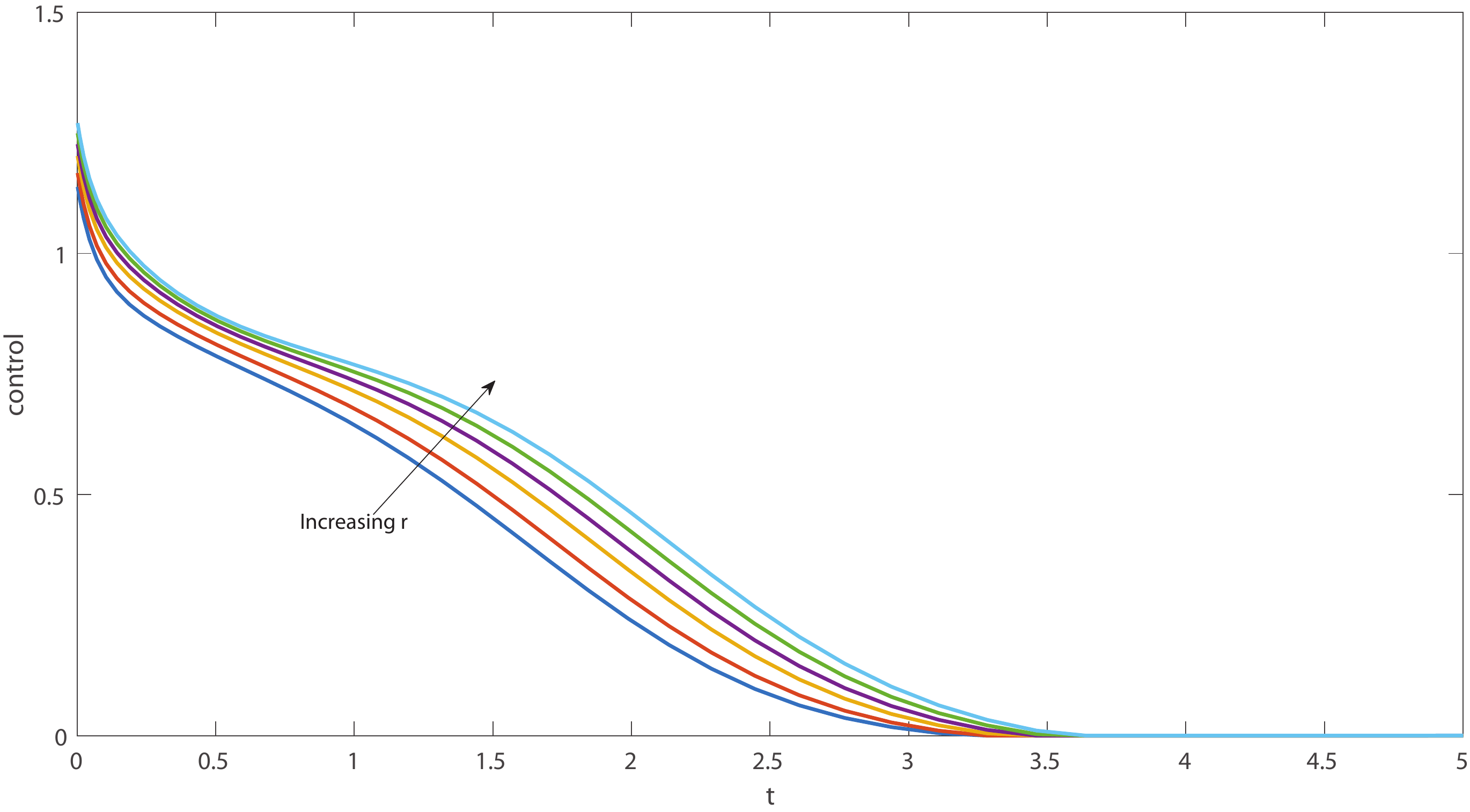}
  }
 \caption{A plot of the optimal controls for increasing $r$ on the interval $[15, 21]$. \label{Fig:optcontrolincr}
 }
 \end{figure}

In Fig.~\ref{Fig:optcontrolincr} we examine the influence of increasing the value of the dimensionless variable $r$.  As $r$ increases we see that optimal control is maintains the same overall shape.  In particular, the initial introduction remains the same, however, the initial drop in the introduction rate is less severe in the case of higher reproductive rates.

%
%\end{document}

\subsection{Stochastic Model $\&$ Sensitivity Analysis}

There are numerous environmental influences that may cause perturbations to the birth and death rates of the invasive species.  Here, we assume that $\beta$ and $\delta$ fluctuate around average values \cite{Gray2011, Zhang2018}. Therefore, the birth and death rates may be treated as random variables such that $\beta \mapsto \beta + \sigma_{\beta} \dot{W}_{\beta}$ and $\delta \mapsto \delta + \sigma_{\delta}\dot{W}_{\delta}$ \cite{Zhang2018}, namely,
\begin{eqnarray}
\label{Stochfeq1} df &=& \left( \beta L \lp \frac{f}{\at}-1 \rp \lp \frac{m}{m+s} \rp fm - \delta f\right)dt \\
\nonumber& & + \sigma_{\beta} \lp \frac{f}{\at}-1 \rp \lp \frac{m}{m+s} \rp fm dW_{\beta} - \sigma_{\delta} f dW_{\delta} \\
\label{Stochmeq1} dm &=& \left( \beta \frac{Lf}{m+s} \lp \frac{f}{\at}-1 \rp \lp m^2 + 2 s^2 \rp - \delta m\right) dt \\
\nonumber& &+ \sigma_{\beta} \frac{Lf}{m+s} \lp \frac{f}{\at}-1 \rp \lp m^2 + 2 s^2 \rp dW_{\beta} - \sigma_{\delta} m dW_{\delta} \\
\label{Stochseq1} ds &=& (\mu - \delta s)dt - \sigma_{\delta} s dW_{\delta}
\end{eqnarray}
where $W_{\beta}$ and $W_{\delta}$ are independent Brownian motions with intensities $\sigma_{\beta}$ and $\sigma_{\delta}$, respectively.

\begin{figure}[H]
 \centering{
  \includegraphics[scale=0.5]{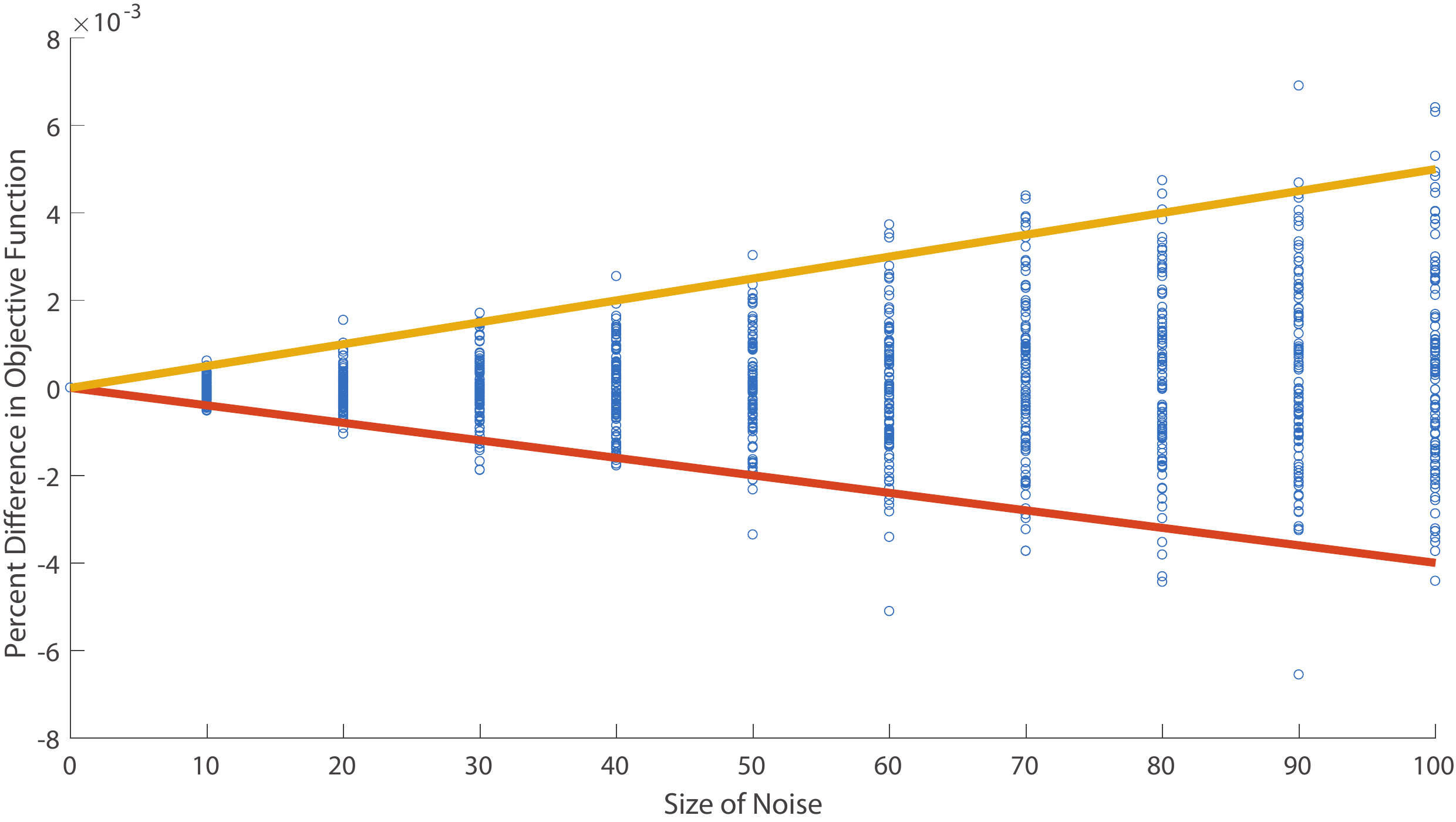}
  }
 \caption{A plot of numerical simulations for increasing size of noise ranging from $0\%-100\%$.  A linear regression is shown for the upper (yellow) and lower (red) $95\%$ confidence interval.  The slopes for the UCL and LCL are $.00052$ and $-.00049$, respectively.}
\label{NoisePlot}
 \end{figure}

Here, we examine the influence of noise on the objective value in simulations with the optimal control given Fig.~\ref{Fig:objfuncs}. The noise on the birth, $d\beta$, and death, $d\delta$, rates are taken from a normal distribution. An Euler–Maruyama method is employed to determine a numerical solution to the stochastic model.  In each simulation we determine the objective function value (Eq.~\ref{objectivefunc}) and compare the percent difference between the objective function value determined in case of no noise.  The results for increasing percent noise are shown in Fig.~\ref{NoisePlot}.  A linear regression through the $95\%$ confidence interval is shown as a reference and the intervals are given in the Table \ref{TableConfidence}.  Notice, that for large amount of noise in the death and birth rates that the objective function is no more than $8\%$ away from the no noise situation.  The slopes of the upper and lower $95\%$ confidence intervals are relatively small and, subsequently, means that the value of objective function is changed only slightly for perturbations in the birth and death rates.  This provides experimental evidence that the determined optimal introduction rate is fairly robust to noise the death and birth rates.

\begin{table}
\begin{center}
 \begin{tabular}{||c c c c||}
 \hline
 Percent Noise & LCL & Mean & UCL \\ [0.5ex]
 \hline\hline
 10 & 0.999914857 &	0.999959708	& 1.000004558 \\
 \hline
 20 & 0.999896955 & 0.999979216 & 1.000061477 \\
 \hline
 30 & 0.999835802 & 0.999986888 & 1.000137975 \\
 \hline
 40 & 0.999776235 & 0.999966538 & 1.00015684 \\
 \hline
 50 & 0.999770905 & 1.000003822 & 1.00023674 \\
  \hline
 60 & 0.999666614 & 1.000169234 & 1.000471855 \\
 \hline
 70 & 0.999749978 & 1.000115747 & 1.000481516 \\
 \hline
 80 & 0.999739497 & 1.000118186 & 1.000496874 \\
 \hline
 90 & 0.999570222 & 1.000001015 & 1.000431809\\
  \hline
 100 & 0.999672431 & 1.000142234 & 1.000612037 \\ %[1ex]
 \hline
\end{tabular}
\caption{The lower (LCL) and upper (UCL) $95\%$ confidence interval for the simulations at increasing noise level.  The LCL and UCL are used to create the regression lines in Fig.~\ref{NoisePlot}.}
\label{TableConfidence}
\end{center}
\end{table}

\section{Conclusions and Future Work}

The mathematical analysis and improvement of models of the TYC eradication strategy are essential to understanding the efficacy of the strategy as a control.  This is especially important in light of recent field studies of reproductivity and survivability of introduced Trojan supermale populations of brook trout (\emph{Salvelinus fontinalis}), in the Big Lost River basin in south-central Idaho \cite{Schill18}.  In this paper, a modification to the classical model of the TYC strategy is proposed that includes intraspecies competition between males and supermales for female mates and a strong Allee effect in the female population.  It is shown that the dynamical system exhibits an extinction and recovery equilibrium solution for realistic parameter values of the carrying capacity and birth and death rates.  In such case, the equilibrium solutions are shown to be asymptotically stable and hence the goal of an effective TYC strategy is to \textit{push} the wild-type populations toward the basin of attraction of the extinction state.

An optimal introduction rate of supermales is determined through optimal control theory and was chosen to minimize an objective function that measures the total amount of wild population and introduced supermales. The optimal introduction rate for the classical model was given in \cite{JingJing2019}.  Here, we compare the influence of including the intraspecies competition for mates and the Allee effect on the optimal introduction rate.  It is determined that the optimal introduction rate yields a significantly smaller objective function value as compared to the classical model. This suggests that it is important to include intraspecies competition and the Allee effect to appropriately determine the overall cost of the TYC eradication strategy.   In addition, this indicates that the overall cost of the strategy is smaller than previously predicted \cite{JingJing2019}. Lastly, a stochastic model is proposed to investigate the influence of noise in the birth and death rates on the objective function's value. These results show the sensitivity in the objective function's value in light of perturbations to the birth and death rates.  Numerical results indicate that the optimal introduction rate is indeed robust to noise in the birth and death rate.  This indicates that the optimal introduction rate is not greatly influenced by  noise in death and birth rates.

\section*{Acknowledgements}

JL and RP would like to acknowledge valuable support from the NSF via DMS-1715377 and DMS-1839993. MB, SB, HC, and TG would like to acknowledge valuable support from the NSF via DMS-1715044.

%\newpage


\begin{thebibliography}{99}

\bibitem{Allee1927}
W.C. Allee, \textit{Animal aggregations}, The Quarterly Review of Biology 2, 367–398 (1927).

\bibitem{Allee1931}
W.C. Allee, \textit{Animal Aggregations. A Study in General Sociology.} Chicago, IL: University of Chicago Press, 1931.

\bibitem{A06}
M. Arim, S. Abades, P. Neill, M. Lima and P. Marquet,
\newblock \emph{Spread Dynamics of invasive species},
\newblock Proceedings of the National Academy of Sciences,
\newblock vol 103, no.2, pp 374-378, 2006.

\bibitem{A12} I. Averill and Y.Lou,
\newblock \emph{On several conjectures from evolution of dispersal},
\newblock , Journal of Biological Dynamics, vol.6, no.2, pp 117-130, 2012.

\bibitem{B07}
C.J. Bampfylde and M.A. Lewis,
\newblock \emph{Biological control through intraguild predation:case studies in pest control, invasive species and range expansion},
\newblock  Bulletin of Mathematical Biology, vol 69, pp 1031-1066, 2007

\bibitem{C01}
J.S. Clark, M. Lewis and L. Horvath,
\newblock \emph{Invasion by extremes: Population spread with variation in dispersal and reproduction},
\newblock The American Naturalist, vol 157, no.5, 2001.

\bibitem{CW07a} S. Cotton and C. Wedekind
\newblock \emph{Control of introduced species using Trojan sex chromosomes},
\newblock Trends in Ecology $\&$ Evolution, vol 22, pp 441-443, 2007.

\bibitem{CW07b} S. Cotton and C. Wedekind,
\newblock \emph{Introduction of Trojan sex chromosomes to boost population growth},
\newblock Journal of Theoretical Biology, vol 249, pp 153-161, 2007.

\bibitem{CW09} S. Cotton and C. Wedekind,
\newblock \emph{Population consequences of environmental sex reversal},
\newblock Conservation Biology, vol 23, pp 196-206, 2009.

\bibitem{Drake2011}
J.M. Drake and A.M. Kramer, \textit{Allee Effects}, Nature Education Knowledge, 3(10):2, 2011.

\bibitem{Gray2011}
A. Gray, D. Greenhalgh, L. Hu, X. Mao, and J. Pan, \textit{A stochastic differential equation SIS epidemic model}, SIAM J. Appl. Math. 71 (2011) 876–902.


\bibitem{GutierrezTeem06} J.B. Gutierrez and J. Teem
\newblock \emph{A model describing the effect of sex-reversed YY fish in an established wild population: the use of a Trojan Y-Chromosome to cause extinction of an introduced exotic species},
\newblock Journal of Theoretical Biology, vol 241, no.22, pp 333-341, 2006.

\bibitem{Gutierrez12} J. B. Gutierrez, M. K. Hurdal, R. D. Parshad and J. L. Teem,
\newblock \emph{ Analysis of the Trojan Y Chromosome Model for Eradication of Invasive Species in a Dendritic Riverine System},
\newblock Journal of Mathematical Biology, vol 64, no.1-2, pp 319-340, 2012.

\bibitem{Schill18} P.A. Kennedy, K.A. Meyer, D.J. Schill, M. R. Campbell, and N.V. Vue,
\newblock \emph{Survival and reproduction success of hatchery YY male brook trout stocked in Idaho streams},
\newblock Transactions of the American Fisheries Society, vol. 147, pp. 419-430, 2018.

\bibitem{Kramer2009}
A.M. Kramer, \textit{The evidence for Allee effects}, Population Ecology 51, 341–354 (2009).

\bibitem{Lenhart_OptControl}
S. Lenhart and J.T. Workman, \emph{Optimal Control Applied to Biological Models}, Chapman $\&$ Hall/CRC, 2007.

\bibitem{L12}  Y. Lou and D. Munther,
\newblock \emph{Dynamics of a three species competition model},
\newblock Discrete $\&$ Continuous Dynamical Systems-A, vol.32, pp 3099-3131, 2012.

\bibitem{JingJing2019}
J. Lyu, P.J. Schofield, K. M. Reaver, M.A. Beauregard, R.D. Parshad, \textit{A comparison of the Trojan Y Chromosome strategy to harvesting models for eradication of non-native species}, Nat. Resour. Model., in press (see arXiv:1810.08279).


\bibitem{M00}
J.H. Myers, D. Simberloff, A.M. Kuris and J.R. Carey,
\newblock \emph{Eradication revisited: dealing with exotic species},
\newblock Trends in Ecology $\&$ Evolution, vol.15, pp. 316-320, 2000.

\bibitem{Allee_1954}
H.T. Odum and W.C. Allee, \emph{A note on thestable point of populations
showing both intraspecies cooperation and disoperation}, Ecol 35: 95-97, 1954.

\bibitem{089} A. Okubo, P.K. Maini, M.H. Williamson and J.D. Murray,
\newblock \emph{The spread of the grey squirrel in Britain},
\newblock Proceedings of the Royal Society of London, Series B, vol.238, pp. 113-125.

\bibitem{P11} R. D. Parshad,
\newblock \emph{On the long time behavior of a PDE model for invasive species control},
\newblock International Journal of Mathematical Analysis, vol 5, no.40, pp 1991-2015, 2011.

\bibitem{ParshadGutierrez11} R. D. Parshad and J. B. Gutierrez,
\newblock{On the Global Attractor of the Trojan Y-Chromosome Model},
\newblock Communications on Pure and Applied Analysis, vol 10, pp 339-359, 2011.

\bibitem{p10} R. D. Parshad and J. B. Gutierrez,
\newblock \emph{On the well posedness of the Trojan Y Chromosome model},
\newblock Boundary Value Problems,  Volume 2010, Article ID 405816, pp 1-29, 2010.

\bibitem{Parshad13} R. D. Parshad, S. Kouachi and J. B. Gutierrez,
\newblock \emph{Global existence and asymptotic behavior of a model for biological control of invasive species via supermale introduction},
\newblock Communications in Mathematical Sciences, vol 11, no.4, pp 951-972, 2013.

\bibitem{P09} N. Perrin,
\newblock \emph{Sex reversal: a fountain of youth for sex chromosomes}?
\newblock Evolution, vol 63, pp. 3043-3049, 2009.

\bibitem{Schill2017} D. J. Schill, K.A. Meyer, and M. J. Hansen,
\newblock \emph{Simulated effects of YY-male stocking and manual suppression for eradicating nonnative brook trout populations},
\newblock North American Journal of Fisheries Management, 37:1054–1066, 2017.

\bibitem{Schill16} D.J. Schill, J.A. Heindel, M.R. Campbell, K.A. Meyer $\&$ E.R. Mamer,
\newblock \emph{Production of a YY Male Brook Trout Broodstock for Potential Eradication of Undesired Brook Trout Populations}.
\newblock North American Journal of Aquaculture, 78(1), 72-83, 2016.

\bibitem{SL15} P. Schofield and W. Loftus,
\newblock \emph{Nonnative fishes in Florida freshwaters: a literature review and synthesis},
\newblock Reviews in Fish Biology and Fisheries, vol. 25, no.1, pp. 117-145.

\bibitem{S97} N. Shigesada and K. Kawasaki,
\newblock ``Biological invasions: Theory and practice",
\newblock Oxford University Press, Oxford, 1997.

\bibitem{Stephens_1999}
P.A. Stephens and W.J. Sutherlan, \emph{Consequences of the Allee effect
for behaviour, ecology and conservation}, Trends Ecol Evol 14: 401-405, 1999.

\bibitem{TGP13} J.L. Teem, J.B. Gutierrez and R.D. Parshad,
\newblock \emph{A comparison of the Trojan Y Chromosome model and daughterless carp eradication strategies},
\newblock Biological Invasions, DOI 10.1007/s10530-013-0475-2, Published online May 2013.

\bibitem{V96}  R. Van Driesche and  T. Bellows
\newblock ``Biological Control",
\newblock  Kluwer Academic Publishers, Massachusetts, 1996.

\bibitem{SDE2013} X. Wang, R. D. Parshad and J. Walton,
\newblock \emph{The Stochastic Trojan Y Chromosome model for eradication of an invasive species},
\newblock Journal of Biological Dynamics, vol 10, issue 1, pp 179-199, 2016.

\bibitem{ODE2013} X. Wang, J. R. Walton, R. D. Parshad, K. Storey and M. Boggess,
\newblock \emph{Analysis of {\it Trojan Y-Chromosome} eradication strategy},
\newblock Journal of Mathematical Biology, vol 68, issue 7, pp. 1731-1756, 2014.

\bibitem{Zhang2018}
W. Zhang and X. Men, \textit{Stochastic analysis of a novel nonautonomous periodic SIRI
epidemic system with random disturbances}, Physica A, 492 (2018), 1290-1301.

\bibitem{Z12} X. Zhao, B. Liu and D. Ning,
\newblock \emph{Existence of global attractor for the Trojan Y Chromosome model},
\newblock Electronic Journal of Qualitative Theory of Differential Equations, No.36, 16, 2012.

\end{thebibliography}
\end{document}